\documentclass[conference]{sig-alternate-2013}
\usepackage{graphicx} 
\usepackage{color} 
\usepackage{amssymb}
\usepackage{epsfig}
\usepackage{url}

\newfont{\mycrnotice}{ptmr8t at 7pt}
\newfont{\myconfname}{ptmri8t at 7pt}
\let\crnotice\mycrnotice%
\let\confname\myconfname%

\permission{Permission to make digital or hard copies of all or part of this work for personal or classroom use is granted without fee provided that copies are not made or distributed for profit or commercial advantage and that copies bear this notice and the full citation on the first page. Copyrights for components of this work owned by others than ACM must be honored. Abstracting with credit is permitted. To copy otherwise, or republish, to post on servers or to redistribute to lists, requires prior specific permission and/or a fee. Request permissions from permissions@acm.org.}
\conferenceinfo{LCDNet'13,}{September 30 2013, Miami, Florida, USA}
\copyrightetc{Copyright 2013 ACM \the\acmcopyr}
\crdata{978-1-4503-2365-9/13/09\ ...\$15.00.\\
http://dx.doi.org/10.1145/2502880.2502887}

\begin{document}

\title{On the Trade-off Between Spectrum Efficiency with Dedicated Access and Short End-to-End Transmission Delays with Random Access in DVB-RCS2}

\numberofauthors{5}

\author{
\alignauthor
Nicolas Kuhn\\ 
       \affaddr{University of Toulouse, ISAE}\\
       \affaddr{National ICT Australia}\\
       \email{nicolas.kuhn@isae.fr}
\alignauthor
Huyen-Chi Bui\\
       \affaddr{University of Toulouse, ISAE}\\
       \email{huyen-chi.bui@isae.fr}
\alignauthor 
J\'{e}r\^{o}me Lacan\\
       \affaddr{University of Toulouse, ISAE}\\
       \email{jerome.lacan@isae.fr}
\and
\alignauthor 
Jos\'{e} Radzik\\
       \affaddr{University of Toulouse, ISAE}\\
       \email{jose.radzik@isae.fr}
\alignauthor 
Emmanuel Lochin\\
       \affaddr{University of Toulouse, ISAE}\\
       \email{emmanuel.lochin@isae.fr}
}

\maketitle

\begin{abstract}

This paper analyses the performance of TCP over random and dedicated access methods in the context of DVB-RCS2. Random access methods introduce a lower connection delay compared to dedicated methods. We investigate the potential to improve the performance of short flows in regards to transmission delay, over random access methods for DVB-RCS2 that is currently under development. Our simulation experiments show that the transmission of the first ten IP datagrams of each TCP flow can be 500\,ms faster with random access than with dedicated access making the former of interest to carry Internet traffic. Such methods, however, are less efficient in regards to bandwidth usage than dedicated access mecanisms and less reliable in overloaded network conditions. Two aspects of channel usage optimization can be distinguished: reducing the duration of ressource utilization with random access methods, or increasing the spectrum efficiency with dedicated access methods. This article argues that service providers may let low-cost users exploit the DVB-RCS2 to browse the web by introducing different services, which choice is based on the channel access method. 


\end{abstract}

\sloppy{
\category{C.2.5}{Computer-Communication Networks}{Local and Wide-Area Networks}
}
\keywords{DVB-RCS2; TCP; Channel access}

\section{Introduction}
\label{sec:introduction}

Satellite links provide a wide range of services for civil and military applications. These services include mobile data transfer, localization, satellite television and Internet web traffic. The Second Generation DVB Interactive Satellite Services (DVB-RCS2) \cite{dvb_RCS2_norm_126}, has been recently adopted as a new standard for satellite communications and is specifically designed for Internet based services. In~\cite[p.~126, sec.~7.2.5.1.3]{dvb_RCS2_norm_126} the authors advise that ``the ST shall by default not transmit in contention timeslots for traffic, but may do this when explicitly allowed by indication in the Lower Layer Service Descriptor or by other administrative means'' making the random access methods used mostly for log in procedure~\cite[p.~182, sec.~9.2.3]{dvb_RCS2_norm_126} and optionally for traffic. The specifications present the potential for introducing both random and dedicated access methods without explaining in which proportion the timeslots of the frames must be divided between them, even though section~\cite[p.~126, sec.~7.2.5.1.3]{dvb_RCS2_norm_126} highlight the preference for dedicated access methods.

Considering that (1) there is a connection delay introduced by dedicated access methods, (2) there is no such delay with random access methods and (3) dedicated access methods enable a more efficient use of the channel capacity, introducing Internet based service in DVB-RCS2 requires adequate measurements on the impact of the access method on the performance of TCP. 

On top of existing services, random access methods could enable the introduction of low cost access to the Internet. Indeed, when slots are not exploited, the servider could warn ``low-cost'' users that they can transmit on these slots without any knowledge on the reliability of the transmission. As a result, by exploiting unused capacity through the implementation of such methods, service providers can propose ``cheap-to-implement'' service, which coverage is important. We propose to evalute, in this paper, the performance that these services could obtain while using random access methods, as compared to dedicated methods.  

The performance of TCP on dedicated access methods has been previously studied in \cite{dama_tcp}\cite{tcp_perf_emulation}. Concerning random access methods, a recent paper \cite{random_dvb_rcs} introduces this idea and some preliminary results. However, the paper does not incude an analysis of TCP over random access methods. Finally, none of these studies analyse and compare the impact of dedicated and random access methods on the performance of TCP.

\sloppy{
In this paper, we present a study on the impact of dedicated access and various random access methods (including Contention Resolution Diversity Slotted ALOHA (CRDSA) \cite{ref_crdsa} and Multi-Slot Coded ALOHA (MuSCA) \cite{ref_musca}) on the performance of short TCP flows~\cite{IOR2009}. We compare random and dedicated access methods, without extensively compare the most recent random access methods, such as R-CRDSA~\cite{ref_RCRDSA}. We propose to discuss the spectrum efficiency of dedicated access methods by presenting the shorter transmision delays enabled by random access methods.
}

We implement a module in NS-2 to model the network component. Note that older DVB-S2/RCS NS-2 module have been proposed in~\cite{ns2_laas,ns2_aberdeen}, which does not implement various random access methods and does not enable to introduce important modifications on the components characteristics.

The rest of the paper is organized as follows: we present the DVB-RCS2 context and the NS-2 module we have implemented to drive this study in Section~\ref{sec:implementation}. In Section~\ref{sec:dedicated_vs_random_cumulated}, we present the overall performance to highlight comparison in terms of spectrum efficiency between dedicated and random access methods considering a varying number of TCP sessions between the gateway and the home user. We compare the transmission times of the first datagrams of a connection in Section~\ref{sec:transmission_time_short}. We evaluate the impact of errors that are introduced with random access methods in Section~\ref{sec:error_short_flows}. Finally we conclude and propose future work in Section~\ref{sec:conclusion}.

\begin{figure}[h]
    \begin{center}
        \includegraphics[width=1\linewidth]{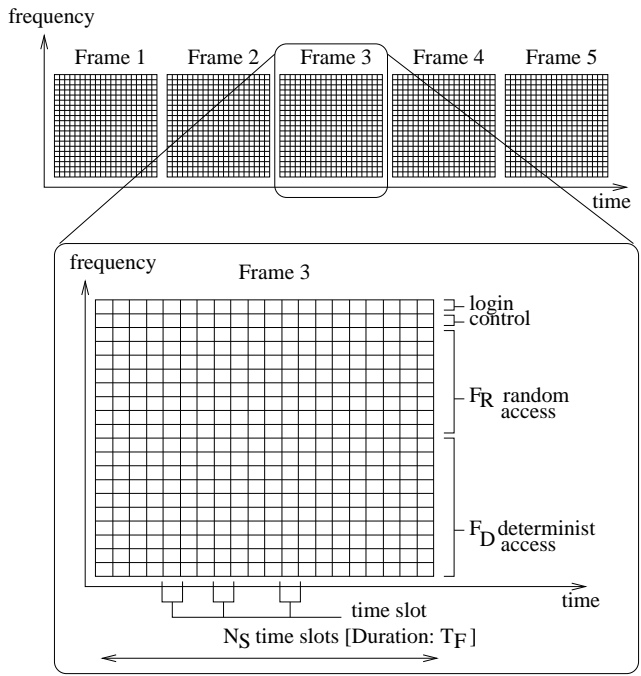}
        \caption{Frame structure}
        \label{fig:frame_struct}
    \end{center}
\end{figure}

\section{DVB-RCS2: access methods and implementation in NS-2}
\label{sec:implementation}

In this section, we describe the network component taking part in the DVB-RCS network, provide details on the access methods and on the implementation in NS-2


\subsection{NCC component and frame architectures}
\label{subsec:ncc_frame}

As illustrated in Figure~\ref{fig:frame_struct}, The Burst Time Plan (BTP) is distributed every $T_{F}=45$\,ms by the Network Control Center (NCC) to the different users that connect to the satellite gateway to indicate them when and how transmit the data. The resource distribution depends on the nature of the flows, the network load and the access methods (random or dedicated). The NCC adapts the repartition of the available slots at each frame, in order to (1) accept late-comer flows; (2) adapt the time slots reservation depending on each user characteristics (different priority between the users); (3) adjust the distribution of time slots depending on the network load; (4) optimize the modcod (modulation and coding) at the physical layer for dedicated access methods. In our scheme, we consider that there is no priority between flows. 

A Multi-frequency time division multiple access (MF-TDMA) scheme where users' packets are distributed over 100 carriers (i.e., 100 frequencies). Thus, frame of 45\,ms length is composed by $100\times 40$ slots. Each slot of length $1.09$\,ms carries $536$ symbols. In the simulations, we consider a "clear sky'' scenario with a Signal to Noise Ratio (SNR) equal to $8.6$\,dB. The slots contain a fixed number of symbols. The number of useful bits transmitted within a slot is determined by the modcod which depends on the channel, the ST characteristics and the access method.

\subsection{Channel access methods}
\label{subsec:access_method}

In this section, we compare dedicated and random access methods.

\subsubsection{Dedicated access methods}
\label{subsubsec:dedicated_access_methods}

We use the standard dedicated access scheme. To ensure the connection, each terminal sends a resource reservation request. This introduces a lower bound to the delay necessary to establish the connection. 

In our implementation (further details are given in \ref{subsec:implementation_details}), the allocation of time slots with this method is divided into three main steps. Firstly, we reserve one time slot for each flow that transmitted data on the previous frame. For one given application, as soon as the connection is established and while there are datagrams of this application in the ST, there will be at least one slot under reservation for the next frame. Secondly, we check in the ST if there are new first IP datagrams from a new flow and allocate them one time slot if possible. Finally, we allocate the slots depending on the number of slots that are free and the maximum capacity that is allowed for each user.

In the conditions described in the previous section ($SNR = 8.6$\,dB), we assume that the users apply a code of rate $R = 2/3$ combined with 8PSK modulation to encode a packet of 920 information bits into  a codeword of 1380 bits, \textit{i.e.} 460 symbols. Due to the encapsulation at the physical layer level, the physical layer data unit is increased to 536 symbols. 

It should be noted that, since an ST does not transmit at a given time on different frequencies, the impact on the number of time slots used in one frame can not be neglected. For the dedicated access methods, in the context of DVB-RCS2 with the parameters given previously, the maximum number of time slots used by a given ST per frame is 40, even if all time slots in the frame are available.

\subsubsection{Random access methods}
\label{subsubsec:random_access_methods}
The random access method implemented in DVB-RCS2 ~\cite{dvb_RCS2_norm_126} is Contention Resolution Diversity Slotted ALOHA (CRDSA)\cite{ref_crdsa}. In~\cite{guidelines_advanced_random_access}, the authors define guidelines to design Random Access methods, and assess the performance of CRDSA. We considered these guidelines to evaluate the improvements provided by Multi-Slots Coded ALOHA (MUSCA), that authorizes  more users to transmit on a given number of time slots~\cite{ref_musca}. 

We evaluate CRDSA (included in the DVB-RCS2 standard~\cite{dvb_RCS2_norm_126}) and MuSCA~\cite{ref_musca} which can be considered as an generalization of CRDSA. 

For the random access, 
data will be transmitted right after the connection with the ST. In order to recover the data, 100 slots are grouped in a random access (RA) block. 
Then, we consider that slots of 2.5 frequencies are linked by the same scheme and that a user can transmit data on each available RA block. 

\begin{figure}[h]
    \begin{center}
	\includegraphics[width=1\linewidth]{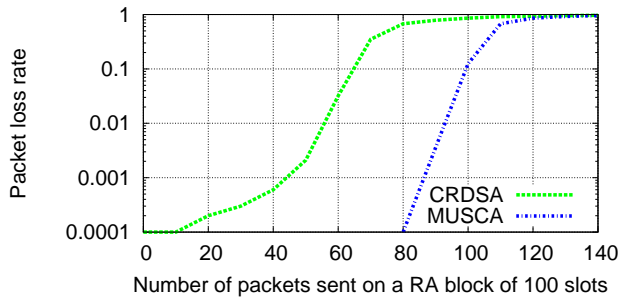}
	\caption{Packet loss rate for random access methods at 5dB}
	\label{fig:perf_crdsa_musca}
    \end{center}
\end{figure}

Users connecting to the satellite with random access generally use an operating point lower than for the dedicated access. 
In the rest of this article, we take a margin of $3.5$ dB. Thus, for the  scenario ``clear sky'', we consider that: $(E_s/N_0)_{random}=(E_s/N_0)_{dedicated}-3.5=8.5-3.5=5$\,dB.

In system using CRDSA, at 5 dB, each user can apply error-correcting code of rate $R_{CRDSA} = 2/3$, associated with QPSK simulation to encode a packet of 613 information bits into a codeword of $920$ bits, \textit{i.e.} 460 symbols. In our simulations, the error-correcting code used is a turbo code. As detailed in~\cite{ref_crdsa}, $N_b$ bursts of length about $530$ symbols are then created. The number of generated bursts depends on the version of CRDSA. In this paper, we study the performance of regular CRDSA-3 ($N_b = 3$). The $N_b$ bursts are transmitted randomly into $N_b$ slots of an RA block.

In the case where the random access method used is MuSCA, users  encode a packet of $680$ information bits with the turbo code of rate $1/4$  associated with QPSK modulation to created codewords of 1380 symbols. The codeword is split into 3 parts to generate 3 bursts (detailed in \cite{ref_musca}) sent on time slots of the same RA block. 

Figure~\ref{fig:perf_crdsa_musca} depicts the performance in terms of packet loss ratio (PLR) depending on the number of packets transmitted per RA block by CRDSA-3 and MuSCA-3. We present specific parameters for both random access methods but do not pretend to lead extensive study on the comparison between their performance. 

The limitations introduced by the ST equipment for the random access methods also need to be considered. As the ST can not transmit on two different frequencies at the same time, the performance of the random access method is limited by the fact that each random access transmit data on different frequencies, for each RA block. The maximum number of packets that a random access method can transmit is given by the ratio between the number of slots per frequencies and by the number of bursts that a given random access method transmit on an RA block.
As an example, if there are 40 time slots per frequency and if the access method sends $3$ bursts per packet (see below), a ST transmits at most $\lfloor 40/3 \rfloor =13$ packets per frame. Note that the different packets are sent on different RA blocks. 

\subsection{Implementation details}
\label{subsec:implementation_details}
We introduce the NCC component in NS-2. 
We adapt the transmission time of IP datagrams by implementing the \texttt{enque()} and \texttt{deque()} methods. 

Each IP datagram is defined by a set of parameters (\textit{e.g.}, application ID, access method, size or remaining bytes). The time at which a datagram reaching the queue of the gateway at the start of frame $i$ will start being transmitted depends on its application ID. We introduce a timer of 3\,seconds to cut the connection when there are no new packet from the application. If this application is new or the timer for this application expired, the next datagram only starts to be transmitted at frame $i+c$ (where $c$ is the number of frames necessary for connection establishment (dedicated access: $c=RTT+1$; random access: $c=1$)). If the communication has already been established and the timer did not expire (\textit{i.e.}, there are less than 3 seconds between the last datagram sent and the new one), the next datagram can be transmitted from frame $i+1$.  

We introduce an other timer that expires every 45 ms and reallocate the bits for the next frame depending on the IP datagrams waiting in the queue and the access method involved. With each frame, we decrement the number of remaining bytes that the ST still has to send. Depending on the random access method, the number of packet sent on an RA block and the error probability (cf Figure~\ref{fig:perf_crdsa_musca}), we randomly determine the erasure state of each packet in the queue. When there are no more remaining bytes for a given IP datagram, it is considered received by the other end at the NS-2 level. More details can be found in~\cite{imple_detail}.

\subsection{Comparison between dedicated and random access methods}
\label{subsec:comparison_random_dedicated}

With dedicated access method, the channel is reserved to the user and it enables the NCC to chose an optimal modcod. As a result, the use of satellite link capacity is optimized. It follows that 1) the communication is reliable and 2) the throughput is maximal.

With random access methods, there is no step of resource reservation request and thus reduce the delay of access to the link. The modcod can not be optimized, as the channel between the home user and the gateway is not known.

We developed a tool that enables us to compare the impact of these access methods on the performance of transport layer protocols. Eventhough the spectrum efficiency is optimized with dedicated access methods, random access methods enables to reduce the time a session uses the ressource. 

\section{Spectrum efficiency}
\label{sec:dedicated_vs_random_cumulated}

We base the following numerical values on the information presented in Section~\ref{sec:implementation}. There are 4000 slots per frame. We also introduce the antenna limitations which limits the maximum number of slots per TCP sessions to 13 for random access methods and 40 for dedicated access method. Moreover, the capacity is fairly shared between the $N_{U}$ users with dedicated access method.

When $N_{U} \leq 100$, a TCP session can tranmit $40 \times 920=36800$ bits per frame with a dedicated access, or $\lfloor 40/3 \rfloor \times 594=7722$ bits with MuSCA as random access. It seems that dedicated access methods enable each TCP session to transmit more data per frame. However, when the load of the network increases, (1) the maximum number of slots available per TCP session decreases with dedicated access methods, and (2) the error probability with random access increases.

In this section, we consider that an application is a TCP session between a ST and the Internet (through the gateway). The objective of this section is to assess the achievable performance with the different access methods when several TCP sessions are simultaneously opened on the satellite link. 

We consider two nodes in NS-2. The first node, which represents the set of ST, transmits a variable number of TCP sessions to the second node, which represents the gateway,  and implements the module detailed in Section \ref{subsec:implementation_details}. The file size is large enough to ensure data transmission over the complete simulation duration (20 s). The size of the IP datagrams is 1500\,bytes, and the queue at the transmitter level is also large enough to prevent overflooding. We use the Linux implementation of TCP Newreno, with SACK option, as transport layer congestion control. 

We present the performance measured at the transmitter level, without distinctions among the different applications. 

\begin{figure}[h]
    \begin{center}
	\centering
        \includegraphics[width=1\linewidth]{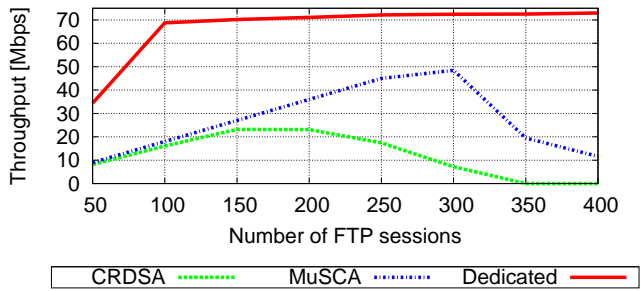}
        \caption{Throughput out of the gateway}
        \label{fig:throughput_ftp}
    \end{center}
\end{figure}

\begin{figure}[h]
    \begin{center}
	\centering
        \includegraphics[width=1\linewidth]{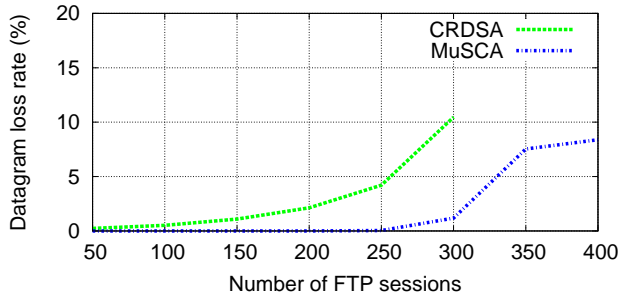}
        \caption{Datagram loss rate}
        \label{fig:error_ftp}
    \end{center}
\end{figure}

In Figure~\ref{fig:throughput_ftp}, we measure the throughput by considering the total number of datagrams transmitted by the ST and the simulation time. Also in Figure~\ref{fig:error_ftp}, we measure the error probability by considering the number of datagrams dropped at the gateway level (\textit{i.e.}, those that can not be recovered by the receiver) and the number of datagrams successfully transmitted. 

If we focus on the performance of random access methods, we can observe they are linked to the type of physical layer codes involved. A trade-off must be found between successfully transmitting data when the load on the network increases and transmitting a larger number of data. Compared to CRDSA, MuSCA enables to transmit more data, even when the load increases. 

When the dedicated access is involved, the cumulated throughput is more important than with any random access methods. Moreover no error is introduced and the network accepts more applications at a given time. 

As a result, we show in this section that the transmission of long files is more efficient with dedicated access methods, as random access methods enable to transmit less data on one given frame and errors might occur. 

We confirm that dedicated access methods are more interesting in terms of spectrum efficiency. This result from the low number of redundancy bits required, as illustrated in Section~\ref{subsec:access_method}. Indeed, the modcod could have adapted during the synchronisation, while more redundancy is required with random access methods.    

Next section looks closer at detailled performance of one given flow in order to explain the interest behind introducing random access methods to carry data traffic.

\section{Transmission time}
\label{sec:transmission_time_short}

Section~\ref{sec:dedicated_vs_random_cumulated} concludes that the transmission of data is more efficient with dedicated access methods. However, considering that (1) there is a connection delay introduced by dedicated access methods, (2) there is an important proportion of short flows in the Internet (measured in~\cite{ciullo09,IOR2009}), we now study the benefits that random access methods can provide in terms of transmission delay for short flows when there are no errors.

\begin{figure}[h!]
	\centering
 	\includegraphics[width=1\linewidth]{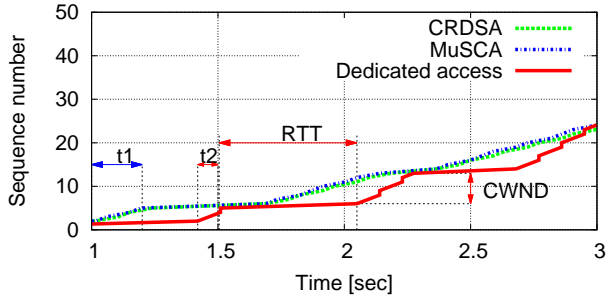}
 	\caption{Evolution of TCP segment sequence number reception }\label{fig:seqno_time}
\end{figure}

In Figure~\ref{fig:seqno_time}, we plot the evolution of the TCP segment sequence decoded at the receiver side when there are 200 TCP sessions. When the number of TCP sessions increases, dedicated access methods allows less capacity for each session, as the bandwidth is fairly shared. The time needed to transmit datagrams with this access is then higher, whereas it is the same with random access methods (the capacity used by each session is not linked to the load on the network). However, with random access, datagrams errors might occur and increase the average transmission time of each datagram. That might require more simulations, and we present here the performance when there is a fixed number of TCP sessions. 

We can see the progression of the congestion window of TCP in the slow start phase, with CWND and RTT presented in the figure. Overall, this figure illustrates that the RTT needed for the connection when dedicated access is involved delayed the transmission of the first datagrams. Also, we can see that the time needed to transmit two datagrams is smaller with dedicated access (denoted $t2$ in the figure) than with random access (denoted $t1$). As a result, with dedicated access, the progression of the congestion window is faster, but starts later. This explains why more datagrams can be transmitted among all the simulation. 

\begin{figure}[h!]
 	\centering
 	\includegraphics[width=1\linewidth]{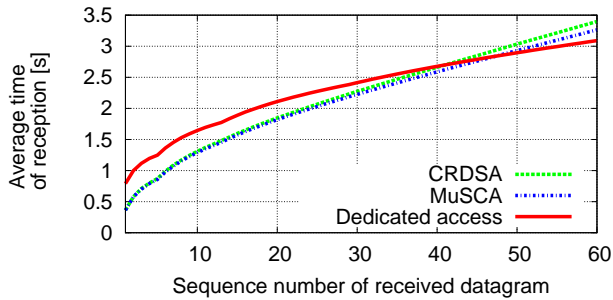}
 	\caption{Cumulated reception time}\label{fig:seqno_time_cumulated}
\end{figure}

In Figure~\ref{fig:seqno_time_cumulated}, we represent the average time needed to transmit a certain number of datagrams for a flow that did not loss datagrams when there are 200 users. The direct connexion with random access methods enables to receive the first datagrams faster. The number of datagrams transmitted before dedicated access enables the reception of the same number of datagrams than random access methods depends on the number of users: when the number of users increases, the transmission of datagrams for dedicated access is slower as the capacity is fairly shared among the sessions. We illustrate that the first datagrams of 1500\,bytes are transmitted quite earlier with random access methods. As an example, at $t=1.5$\,s, 8 datagrams have been received with dedicated access method, instead of 14 with MuSCA. At $t=2.7$\,s, both MuSCA, CRDSA and dedicated access have received 42 datagrams. 

When the load of the network is low, it highlights a clear advantage for short flows (less than 40 datagrams) when using random access methods. The size of the flow from which it is more interesting to use a dedicated access depends on the load of the network. 
This section argues than even if more redundancy bits are required with random access methods, this type of access enables a faster transmission of short flows. Moreover, it could be of interest to transmit this data before the network gets more loaded. 

\section{Errors and waste of bandwith}
\label{sec:error_short_flows}

In thise section, we propose to assess the impact of the errors taht can occur with the random access methods. 

We gather the minimum, median and maximum number of datagrams that have been transmitted in Table~\ref{tab:med_min_max_per_app}. 

\begin{table}[!ht]
        \centering
        \caption{Number of datagrams sent per TCP session} \label{tab:med_min_max_per_app}
                \begin{tabular}{cc|ccc}
                \hline
                \textbf{Access} & \textbf{Number} & \multicolumn{3}{c}{\textbf{Nb datagrams sent}} \\
                \textbf{method} & \textbf{TCP sessions} & \textbf{Min} & \textbf{Med} & \textbf{Max} \\
                \hline
                Dedicated & 100 & \multicolumn{3}{c}{1146} \\
                Dedicated & 200 & \multicolumn{3}{c}{591} \\
                Dedicated & 300 & \multicolumn{3}{c}{400} \\
                Dedicated & 400 & \multicolumn{3}{c}{304} \\
                \hline
                MuSCA & 100 & 300 & 300 & 300 \\
                MuSCA & 200 & 300 & 300 & 300 \\
                MuSCA & 300 & 78 & 296 & 300 \\
                MuSCA & 400 & 1 & 1 & 199 \\
		\hline
                CRDSA & 100 & 204 & 272 & 272 \\
                CRDSA & 200 & 31 & 271 & 271 \\
                CRDSA & 300 & 1 & 190 & 190 \\
                \hline
                \end{tabular}
\end{table}

With the dedicated access, the maximum number of transmitted datagrams decreases due to the fair share of the capacity among the different flows and no difference can be noticed between the minimum, median and maximum number of datagrams transmitted for all simulation cases. As an example, when the number of users increases from 50 to 250, the maximum number of datagrams sent over 20 s for one application decreases from 1180 to 500 datagrams. 

With the random access methods, the minimum and maximum are set at the maximum value authorized by ST limitations and explains the different results detailed in the previous subsection. When error events start to occur, the minimum value greatly decrease, but the median value is stable, as for many TCP sessions, the receiver could recover all the data. However, when the load on the network increases, the median value decreases greatly, but some TCP sessions can still transmit the maximum quantity of data. 

From these results, it appears that the increase of the number of applications impacts in different ways the performance of each application due to errors that occur at different moments of the transmission. Also, errors that occur with random access methods should be avoided, as this is pure waste of bandwidth. 

\section{Conclusion}
\label{sec:conclusion}

Section~\ref{sec:dedicated_vs_random_cumulated} measured the benefits, at the transport layer level, provided by an optimized spectrum effiency with dedicated access methods. In Section~\ref{sec:transmission_time_short}, we limit this advantage by comparing the time necessary to transmit the first ten IP datagrams of each TCP flow, which can be 500\,ms faster with random access than with dedicated access, making the former well suited to carrying web data traffic. However, Section~\ref{sec:error_short_flows} highlights that a control of admission is required to prevent waste of bandwidth.   

Moreover, in ``Why latency matters to mobile blackhaul''\footnote{Published by O3b Networks and Sofrecom. Available at: \url{http://www.o3bnetworks.com/telcos/mobile-backhaul}}, the authors explain that Google measured that ``an additional 500\,ms to compute (a search) [...] resulted in a 25\% drop in the number of searches done by users''. This highlight our interest to increase the transmission of small HTTP requests. Moreover, the introduction of random access methods in DVB-RCS2 would let the service providers propose Internet access which is cheap to set up and worth the integration at the end user level.  

The results presented in the article let us argue that, for a given TCP session, switching from random access to dedicated access is of interest. It increases the end user experiences of the service and the spectrum efficiency, as during connection establishment, the first data packets would already be sent on RA blocks. This idea is presented in IP Over Satellite (IPOS) standards, which are published under the reference: ``TIA 1008-A - IP over Satellite (IPOS)''. We argue for the integration of such strategy in the current DVB-RCS2 standards. 

In order to optimize the transmission of short flows and the spectrum efficiency, we propose a capacity distribution that at $t=k \times T_{F}$ ($T_{F}$ is the frame duration): (1) estimate the load on the network for $t \in [k \times T_{F} ; (k+1) \times T_{F}]$ and determine $SEQ(N)$, the sequence number from which it is interesting to switch from random access method to dedicated access method for each flow; (2) determine the number of RA block, considering the users that paied for a guaranted service, the number of flows for which the sequence number of the datagram to sent is less than $SEQ(N)$ and the load estimation; (3) affect capacity to all flows which current sequence number is greater than $SEQ(N)$ with dedicated access. Such methods has been evaluated in the context of aeronautical communications~\cite{dvb_rcs2_mix_random_dama}.

We propose, as a future work, to investigate on the implementation of this algorithm in our context, which will increase the transmission of short flows while optimizing the spectrum efficiency.

\section*{Acknowledgements}

The authors thank Thales Alenia Space, CNES and NICTA for support and several discussions.

\bibliographystyle{ieeetr}
\bibliography{physical_channel_access.bib}

\end{document}